\documentclass[letterpaper]{article}
\usepackage[preprint]{spconf}

\usepackage{amsmath,epsfig}

\usepackage{subcaption}
\usepackage[utf8]{inputenc}
\usepackage{amsmath,amssymb}

\usepackage{graphicx}
\usepackage{grffile}

\usepackage{color}
\usepackage{array}

\usepackage{algorithm}
\usepackage{algpseudocode}

\usepackage{float}
\restylefloat{table}

\begin{document}\sloppy

\title{Guitar Solos as Networks \footnote{This paper appears in the Proceedings of the 2016 IEEE International Conference on Multimedia and Expo (ICME 2016), IEEE, Seattle, 2016, pp. 1-6.}}
%
\name{Stefano Ferretti\thanks{This paper appears in the Proceedings of the 2016 IEEE International Conference on Multimedia and Expo (ICME 2016), IEEE, Seattle, 2016, pp. 1-6.}}
\address{Department of Computer Science and Engineering, University of Bologna\\
 Mura A. Zamboni 7, I-40127 Bologna, Italy\\
 s.ferretti@unibo.it}


\maketitle

\begin{abstract}
This paper presents an approach to model melodies (and music pieces in general) as networks. Notes of a melody can be seen as nodes of a network that are connected whenever these are played in sequence. This creates a directed graph. By using complex network theory, it is possible to extract some main metrics, typical of networks, that characterize the piece. Using this framework, we provide an analysis on a set of guitar solos performed by main musicians. The results of this study indicate that this model can have an impact on multimedia applications such as music classification, identification, and automatic music generation.
\end{abstract}

\section{Introduction}


Nowadays, there is a common trend in research to model everything as a network.
In particular, complex network theory is a mathematical tool that connects the real world with theoretical research, and is employed in many fields,
ranging from natural and physical sciences to social sciences and humanities 
\cite{newmanHandbook}.
Technological, biological, economic systems, di\-sease pathologies
can be modeled in the same way.
Focusing on (multi)media contents, it has been proved that language, for instance, can be seen 
as a complex network 
\cite{Cong2014598,Grabska}.
In this paper, we show that musical pieces can be treated as complex networks as well.

When dealing with audio, the main concern has been on the issue of digitalizing it 
or to synthesize, represent, reproduce sounds, by employing a variety of sound generation techniques.
Attention has been paid on transmitting, indexing, classifying, clustering, summarizing music 
\cite{Anglade09genreclassification,Patra:2013}.
However, the idea of capturing some general characteristics of a melody (and harmony) is somehow an overlook aspect.
In literature, there are works in the field of computer science that focus on musical scores 
\cite{Prather:1996}, as well as 
works on the automatic transcription of the melody and harmony \cite{Ryynanen:2008}.
As concerns music information retrieval, 
techniques worthy of mention are acoustic-based similarity measures \cite{Berenzweig:2004}, compression-based methods for the classification of pieces of music \cite{Cilibrasi:2004}, statistical analyses and artificial neural networks \cite{Manaris:2005}.
Finally, in \cite{Manaris2005} artificial intelligence is employed to capture statistical proportions of music attributes, such as pitch, duration, melodic and harmonic intervals, etc. 

Studies on music can be based on symbolic data (music scores) or on audio recordings. Symbolic music data eases the analysis in several music application domains.
For example, finding  the  notes  of a melody in an audio file can  be  a  difficult  task, while with symbolic music, notes are  the  starting  point  for  the analysis.  
Thus, in general traditional musicological concepts such as melodic and harmonic structure are easier to investigate in the symbolic domain, and usually more successful \cite{Knopke2011}.

In this paper, we develop a model that allows capturing some essential features of a musical performance (a music track).
We will focus on melodies, and specifically on ``solos'', which are a part of a song where a performer plays (quite often improvises) a melody with accompaniment from the other instruments.
It is quite common in music theory asserting that solos performed by musicians are bound to their technical and artistic skills. Indeed, musicians are recognized for their own ``style'' in playing a solo over a music piece, that identifies a sort of musical ``language'', typical of that musician. 
It is not by chance that an artist can be recognized from others, and that we can classify artists in categories and hierarchies.
The goal of this work is to make a step further toward the identification of the rules and characteristics of the music style of a certain performer. 
If a music line is conceived of as a complex network of musical units (notes, rests) and their relations, it is expected to exhibit emergent properties due to the interactions between such system elements. 
Complex networks provide appropriate modeling for music as a complex system and powerful quantitative measures for capturing the essence of its complexity. 

As a proof of concept, we retrieved and analyzed different solos of some main guitar players.
Namely, the artists are Eric Clapton, David Gilmour, Jimi Hendrix, BB.~King, Eddie Van Halen.
The selection of guitar as instrument and these particular artists is motivated by the fact that there is a quite active community of guitar enthusiasts that share musical scores on the Web. Scores are published and formatted, usually, by employing description schemes that are alternative, easier and more intuitive to read, with respect to the classic musical sheets.
These schemes are based on guitar tablatures, and there is a wide list of software applications and libraries to handle digital representations of such scores.
This simplified the creation of the database.


During the analysis, different measures are calculated, typical of complex network theory. We measure the length of solos, the dimensions of the networks, the degree distribution, distance metrics, clustering coefficient and, finally, we identify that the network representations of certain solos are small worlds. The paper discusses how these metrics are related to the ``style'' of the performer.

The outcomes of such study can have an impact on multimedia applications and on studies of music classification and identification, in general.
While probably a music track cannot be fully described via mathematical measurements, nonetheless, these measures can help in discriminating among the main features of a performer and a music track.
Such results can be employed as building blocks inside media applications for the automatic generation of digital music with certain specific characteristics (e.g., the generation of a solo ``à la'' Miles Davis). Such applications could be extensively exploited in didactic scenarios, automatic music generation applications, and multimedia entertainment.

The reminder of this paper is organized as follows.
Section \ref{sec:model} describes the network model for music solos. 
Section \ref{sec:assessment} discusses on an assessment on a list of guitar solos.
Section \ref{sec:res} presents the obtained results.
Finally, Section \ref{sec:conc} provides some concluding remarks.

\section{A network model for music tracks}
\label{sec:model}


Before dealing with the model, a brief terminology is introduced to avoid potential ambiguities.
A song is a musical piece, which is composed of multiple, simultaneous sounds played by different instruments.
The part of the song played by a single instrument is referred as a track. Notice that an instrument can play different tracks in the song (e.g.~there are multiple instruments of the same type, or the tracks have been overdubbed).


The solo is a part of a track where a performer is playing with unobtrusive accompaniment from the other instruments. 
It is employed quite often in jazz, blues, rock songs, where the solo has, usually, the twofold role of: i) creating a melody (in certain cases an improvised melody), which is alternative to the main melody of the song, and ii) showing off the skills of the performer, due to the technical difficulties to play that solo, or due to the ability of the performer to create an intense, touching melody.
Needless to say, solos are melodies composed of notes (or groups of simultaneous notes). A note can be a pitched sound or a rest, lasting for a certain duration.



We can represent a track as a directed network, whose nodes are the notes played by the performer. When a performer plays a note $x$, followed by a subsequent note $y$, we add the two nodes $x, y$ in the network and a directed link $(x, y)$ from $x$ to $y$. 
If, for instance, the player subsequently plays another note $z$ followed by the note $x$, we add another node $z$ and a link $(z, x)$ leaving from $z$ to the already existing node $x$.
Networks can have cycles, i.e~a performer can play two subsequent notes of the same type.
Weights can be associated to links $(x, y)$, depending on how many times that link $(x, y)$ is present in the network, i.e.~how many times the performer plays a sequence of the two $x, y$ notes in his solo.

%

\begin{figure}
\centering
\begin{subfigure}{\linewidth}
   \centering
  \includegraphics[width=.57\linewidth]{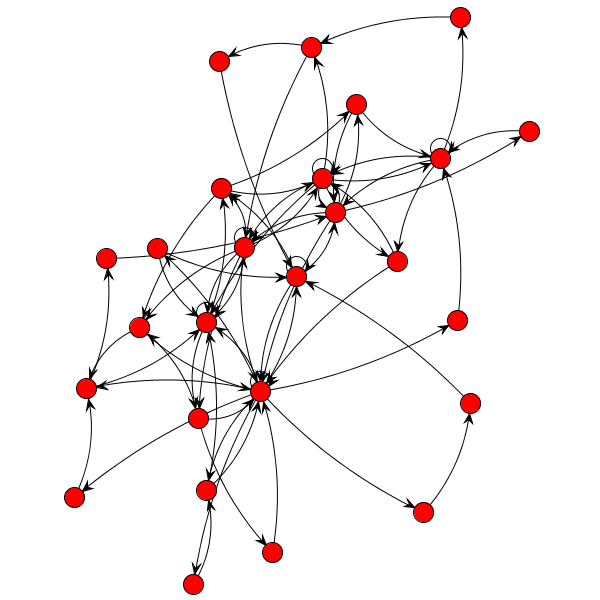}
   \caption{B.B.~King -- Rock me baby}
  \label{fig:bbking}
\end{subfigure}%
\\
\begin{subfigure}{\linewidth}
  \centering
  \includegraphics[width=.57\linewidth]{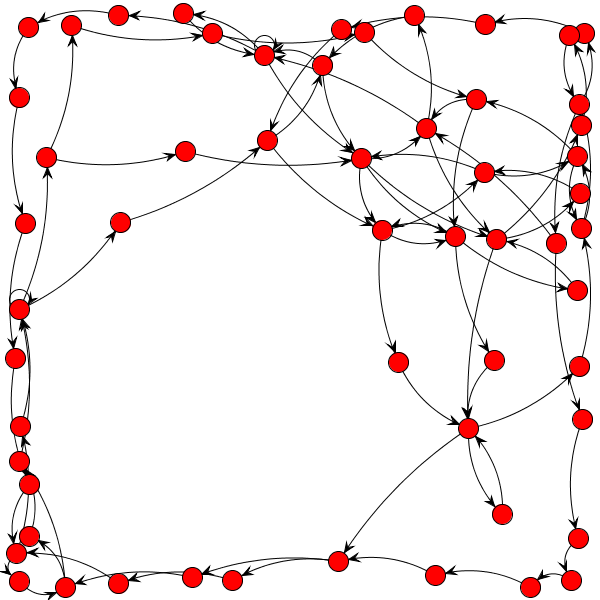}
  \caption{David Gilmour (Pink Floyd) -- Comfortably numb (first solo)}
  \label{fig:confNumb}
\end{subfigure}
\\
\begin{subfigure}{\linewidth}
  \centering
  \includegraphics[width=.57\linewidth]{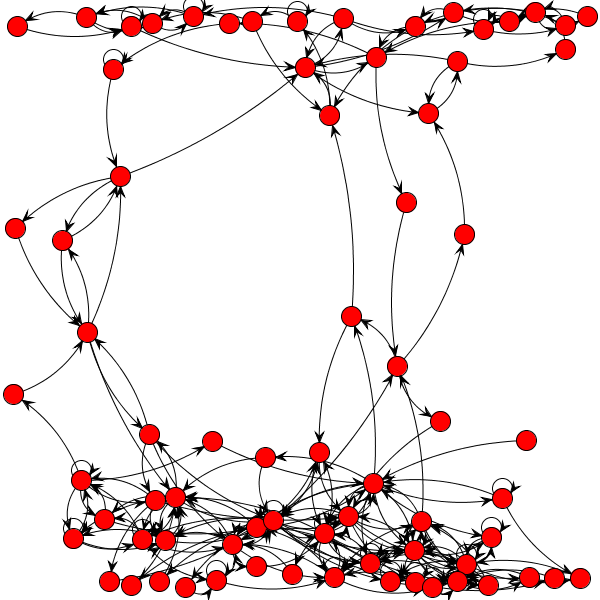}
  \caption{Eric Clapton (Cream) -- Crossroads (second solo)}
  \label{fig:crossroads}
\end{subfigure}%
\\
\begin{subfigure}{\linewidth}
  \centering
   \includegraphics[width=.57\linewidth]{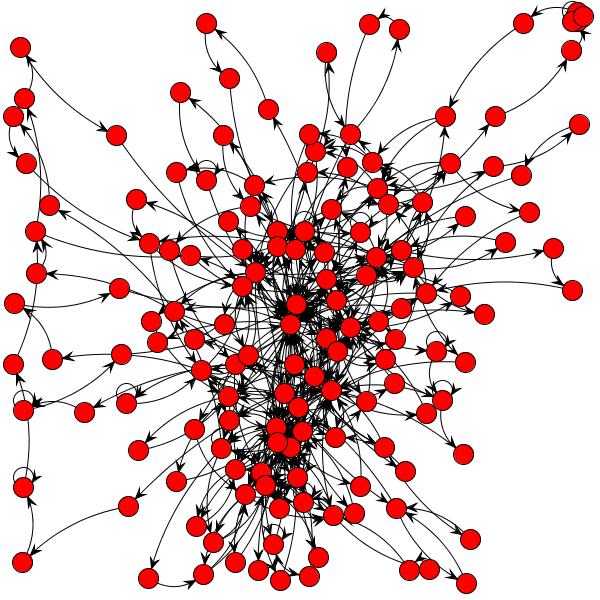}
  \caption{Jimi Hendrix -- Red House}
  \label{fig:redHouse}
\end{subfigure}
\caption{Four examples of networks from different scores}
\label{fig:someNets}
\end{figure}

As concerns the amount of possible nodes in a network, in the occidental music, an octave (the interval between one musical pitch and another with half or double its frequency, which is the same note but lower or higher in pitch) is composed of twelve sounds.
Focusing on electric guitar, as an example, it is usually possible to create sounds belonging to four octaves. 
(Actually, this is a simplified measure, just to give an idea on the amount of possible nodes).
Then, each note has an associated duration. Two notes of the same pitch with different durations are considered as two different network nodes.
Rests and chords are other possible nodes of a network. In fact, it is quite usual to hear performers playing multiple simultaneous notes to create multiple voices in the melody they are composing in the solo.
Therefore, according to this model a guitar solo (and, in general, a track) network can be composed of hundreds of nodes.

Figure \ref{fig:someNets} depicts four examples of networks derived from four famous blues/rock guitar solos. 
It is interesting to observe that these networks are quite different one from the other.
Figure \ref{fig:bbking} shows a simple network, with some nodes that have higher in/out degrees (i.e., number of links entering or leaving a node). Figure \ref{fig:confNumb} has a more linear structure, suggesting that the melodic line was ``simple'', with poor repe\-ti\-tions of single notes.
Figure \ref{fig:crossroads} appears to be a clustered network, with some few nodes connecting the two clusters.
Finally, Figure \ref{fig:redHouse} has a quite complex structure, with many nodes and with the presence of several hubs.
This suggests that it might be interesting to assess if different artists do have different characteristics that, statistically, produce different types of networks.

It is worth noticing that usually a melodic line is strongly influenced by the underlying harmonic, chord progression. Modulations and chords' alterations are quite common in jazz compositions, where improvisations use and outline the harmony as a foundation for melodic ideas.
In this study we do not focus on the harmonic aspects, just taking the melodic line to create the networks.
This can be regarded as a further work.
Actually, the considered songs were all blues/rock ones, without any important modulation in the harmonic progression during the solo.


\section{An assessment on guitar solos}
\label{sec:assessment}

While the study is applicable to all music tracks, regardless on the musical instrument, this study focuses on guitar solos. The reason is quite simple: guitar is probably the most popular musical instrument and there is a vast amount of information available on the Web.
Besides classic musical sheets, alternative notation systems have been devised (e.g.~guitar tablature) and a wide set of software tools is available. 
This allowed creating a wide database of guitar solos. 

A main clarification is that, even if these measures are quantitative, they do not aim at ranking the ability of performers or the ``beauty'' of a solo with respect to another. These are (subjective) opinions which are not under investigation here. 
The aim of these measurements is to extract some main characteristics of performers and their solos, that may serve to perform classifications, build applications for the automatic matching, identification or even automatic generation of media compositions respecting some musical genre or style.


A database was created by downloading guitar scores available in several dedicated Web sites (e.g.~A-Z Guitar Tabs, 
The Ultimate Guitar Tabs).
It is important noting that these scores are typically cooperatively provided by users; thus, these sources might have some minor errors (that nevertheless, do not alter the general scope of this study). 
Retrieved scores were available in Guitar Pro or Power Tab formats. 
Through the use of an existing python library, named pyGuitar 
(http://pyguitarpro.readthedocs.org),
these scores have been manipulated so as to isolate the solo guitar part, extract the guitar solo and export it in a musicXML format. 
Such a process was
(mostly) automatic.
MusicXML is a standard open format for exchanging digital sheet music, that can be easily managed my means of an XML parser.

%
Solos were isolated from the rest of the track, when usually the guitar plays a rhythmic role.
The idea was to extrapolate the melodic line played by the guitar from the rest of the song. 
In certain cases, the track was completely instrumental with the melodic line played by the performer, entirely. In other cases, the performer was continuously playing melodic lines during the song, in alternation with the main melodic line (i.e.~voice). 
In both these cases, a supervised process was performed in order to isolate the ``solo'' parts.

We consider aggregate measures involving the whole solos, that have different durations. 
The different durations of solos have an impact on some of the considered quantitative measures (e.g.~amount of played notes and number of nodes in the networks). 
This allows obtaining a general view of the solo, that considers its duration as a main feature. 
An alternative option might be to normalize these values by dividing them over the duration of the solo. 

The selection of artists was based on the ''importance`` of the performer and the amount of songs available in the Web for that performer.
The idea was to consider a wide range of different musical styles, to see if there are any differences in the corresponding obtained networks. 
We thus selected those musicians that, according to the general music criticism, have a unique playing style. 
Briefly, the artists are: 
i) \textit{Eric Clapton}, a rock-blues guitar legend (the slogan ``Clapton is God'' testifies this); 
ii) \textit{David Gilmour}, singer and guitarist of the famous Pink Floyd rock group, known for his melodic and intense guitar solos; 
iii) \textit{Jimi Hendrix}, which is considered the most important electric guitar player of all times; 
iv) \textit{B.B.~King}, a blues master, quite often referred as ``The King of the Blues'' that inspired several generations of artists; 
v) \textit{Eddie Van Halen}, which is considered one of the most influential hard-rock guitarists of the 20th century.

\section{Results}
\label{sec:res}

In this section, we report some aggregate measures for the considered performers. 
In each figure, concerning a given metrics, a column refers to a performer. For each performer, results for his considered tracks are reported as dots in the column, together with the mean value and standard deviation.

\subsection{Length of solos, number of nodes}

\begin{figure}[htbp]
   \centering
   \includegraphics[width=.8\linewidth]{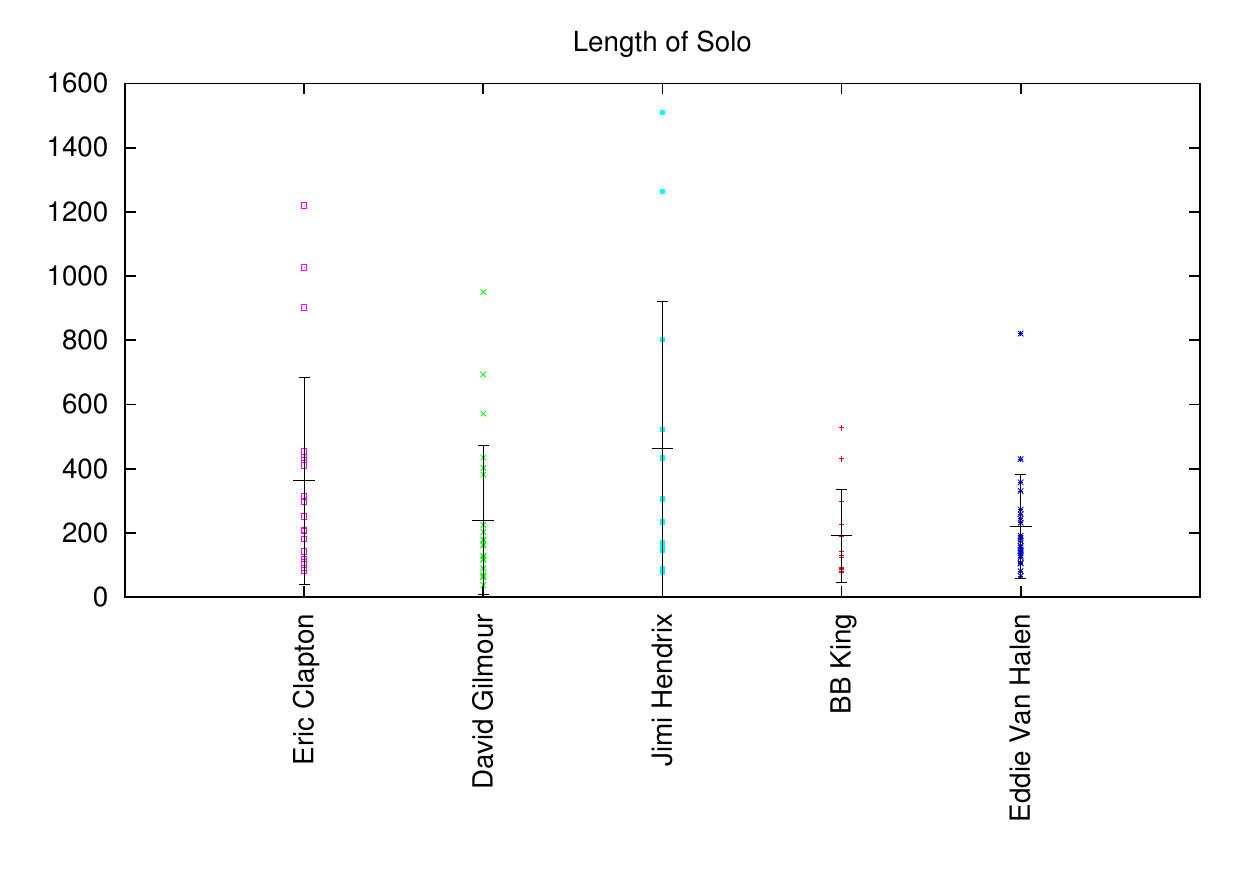}
   \vspace{-.5cm}
   \caption{Length of solos}
   \label{fig:soloLength}
   \vspace{-.3cm}
\end{figure}

Figure \ref{fig:soloLength} shows the length of solos for different tracks, measured as the amount of notes or rests composing the solo. 
This metric is different to the duration time of the solo, even if they are related. In fact, during a bar, one might play a fast sequence of notes with short duration (e.g.~a set of $16$ sixteenth notes) that would occupy the same temporal time of a single whole note.

This metric might express how much a performer is inclined to elaborate the melodic line he is creating during the solo. But it is also strongly related to the music genre. For instance, solos in jazz compositions are usually quite longer than modern pop-rock ones.

\begin{figure}[htbp]
   \centering
   \includegraphics[width=.8\linewidth]{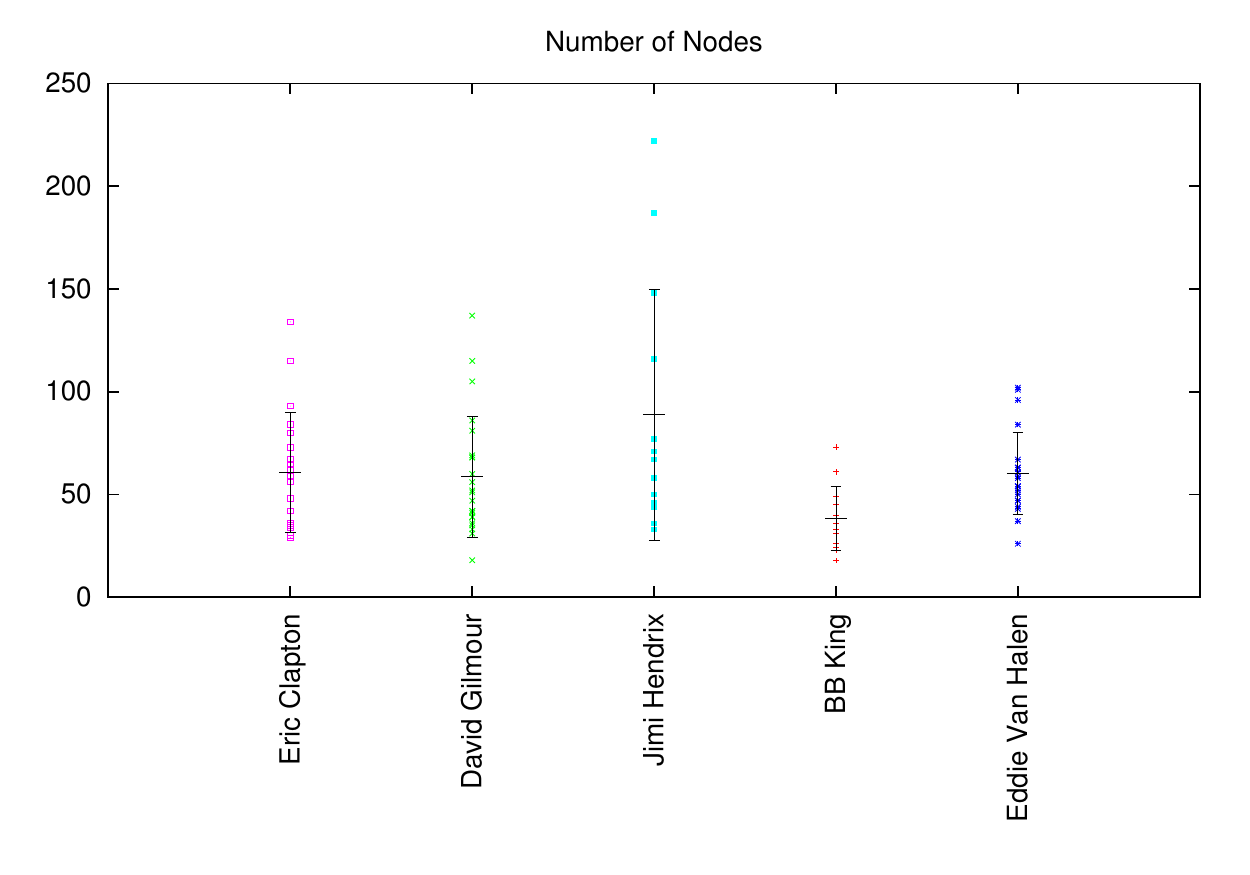}
   \vspace{-.5cm}
   \caption{Number of Nodes}
   \label{fig:numNodes}
      \vspace{-.5cm}
\end{figure}

The metric above can be analyzed together with the number of nodes, reported in Figure \ref{fig:numNodes}, that measures the number of different notes played during a solo.
Results suggest that Hendrix has a richer vocabulary with respect to others (in the considered database, he has longer solos and a higher average number of nodes). 
Moreover, they capture perfectly the style of B.B.~King, noted for his simpler (yet ``touching'') solos. In fact, his solos have a lower average amount of nodes (and lower lengths), with respect to others.

\subsection{Degree distribution}

\begin{figure}[htbp]
   \centering
   \begin{subfigure}{\linewidth}   \centering
   \includegraphics[width=.8\linewidth]{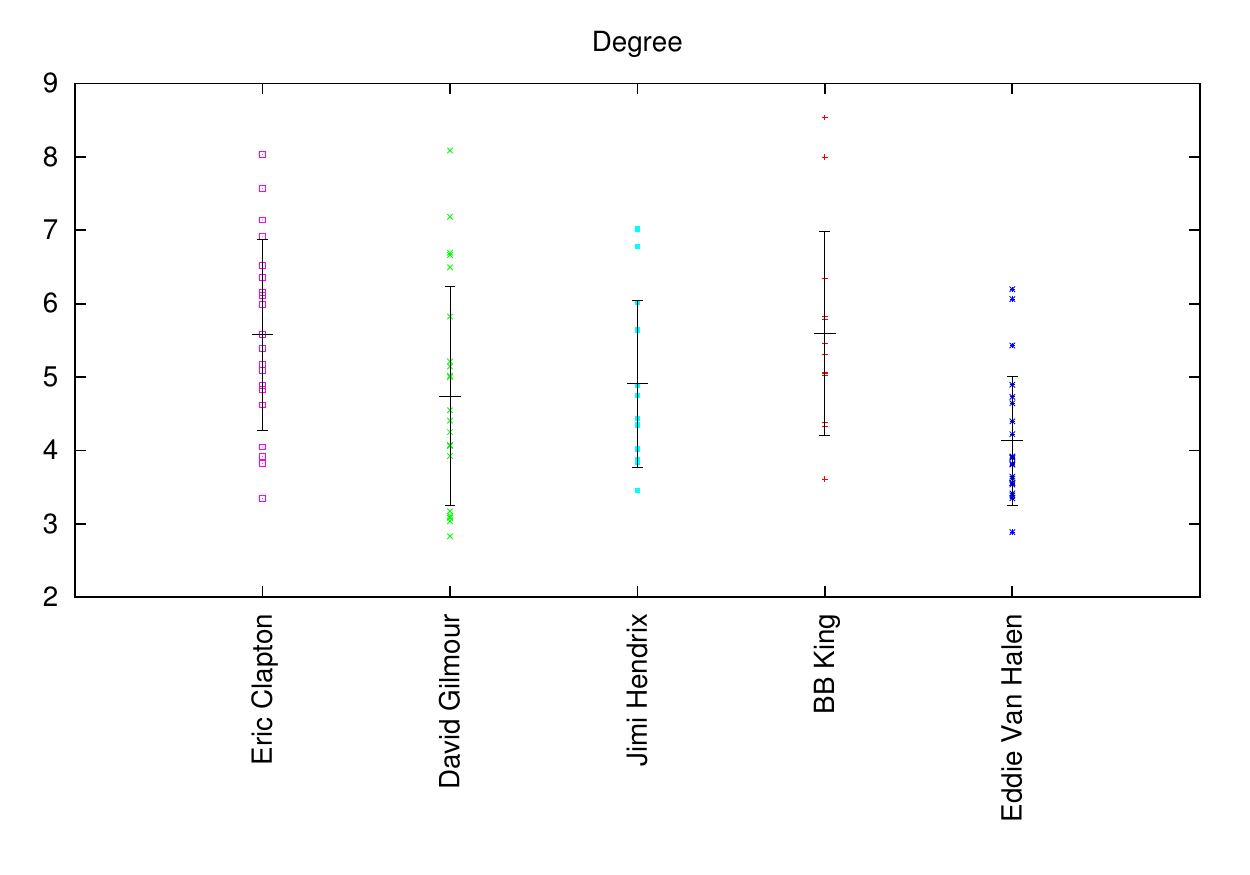}
   \end{subfigure}
   \begin{subfigure}{\linewidth}   \centering
   \includegraphics[width=.8\textwidth]{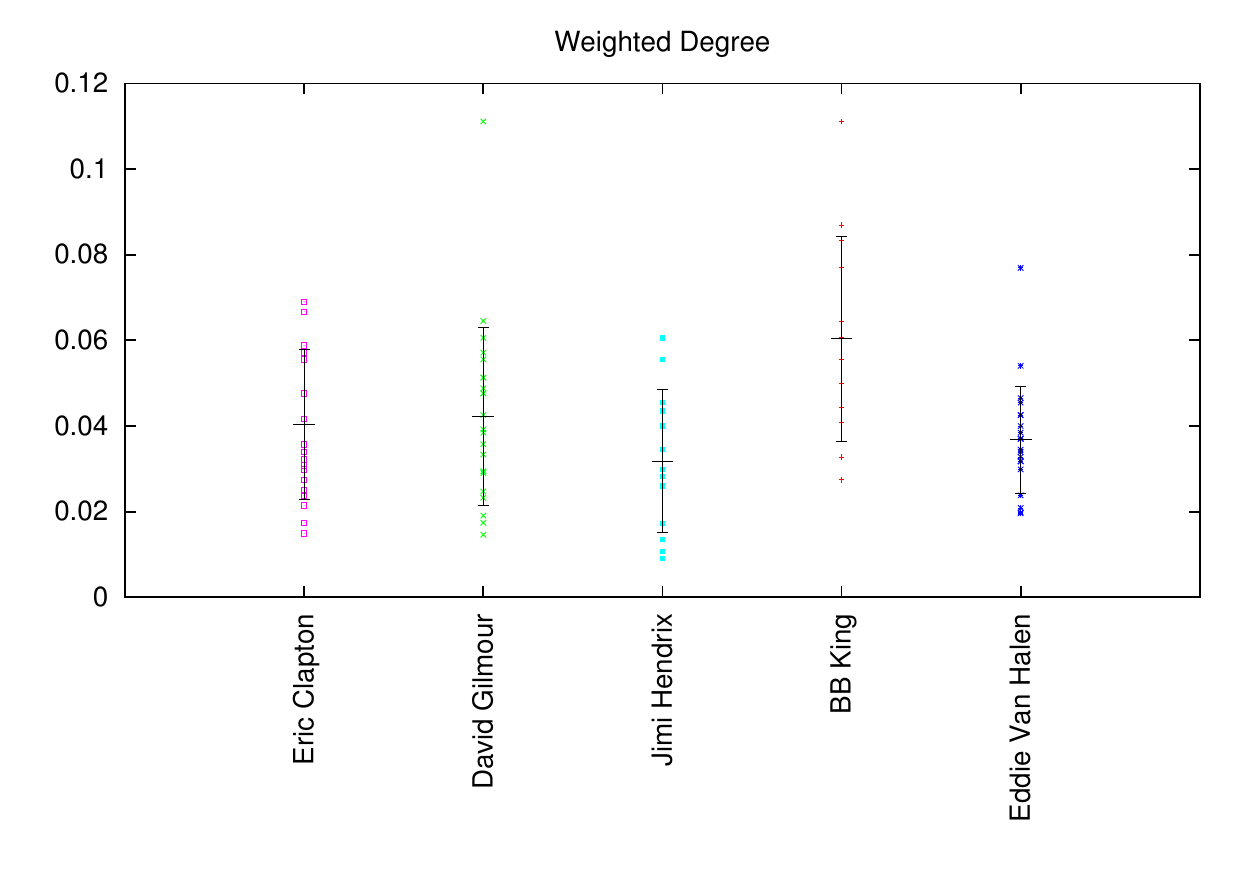}
   \end{subfigure}
   \vspace{-.5cm}
   \caption{Degree distributions}
   \label{fig:degree}
   \vspace{-.5cm}
\end{figure}

Figure \ref{fig:degree} reports the average degrees of solos of the considered performers. In particular, degrees 
and weighted degrees are shown.
Specifically, the degree of a node $n$ is the amount of links that connect $n$ with other nodes in the network (included $n$ itself, in this case a loop would be performed). The degree counts how many times the performer decides playing a note, after (and before) playing another one.
Since the graph is directed, one should consider not only degrees, but also the in-degrees (number of links arriving at a certain node) and out-degrees (number of links leaving a certain node); the degree of a node is the summation between its in-degree and out-degree. In this study, in-degrees and out-degrees have been measured but they are not shown since trends are similar to those obtained for degrees.


As concerns weighted degrees, a weight was assigned to each link, measuring the amount of times the solo ``traverses'' that link (i.e., how many times the performer played those two notes in sequence). Thus, the weighted degree is the sum of the weights of links of a given node.

These measures provide slightly different results. In general, we can say that Eddie Van Halen has lower average degrees, testifying his inclination to create various and complex solos. Again, results confirm that B.B.~King was used playing and repeating specific combinations of notes.



\subsection{Distances}


The average distance is the average shortest path length in a network, i.e.~the average number of steps along the shortest paths for all possible pairs of nodes.
The distance between two nodes $x, y$ is the shortest sequence of notes played in a solo starting with $x$ and ending with $y$.
It is worth noting that when considering distances in most directed networks obtained from the considered solos, the average path length is infinite, since such directed networks are not strongly connected, i.e.~in these networks it is not possible to create a path from a node to another, for all possible pair of nodes, by employing their directed links.
Thus, we report the distance by considering the corresponding undirected networks (i.e.~removing the direction information and considering links as bidirectional ones).
In this application domain, a higher distance reveals a higher complexity of the solo, in terms of amount of notes and how these are connected during the solo construction.

\begin{figure}[htbp]
   \centering
   \includegraphics[width=.8\linewidth]{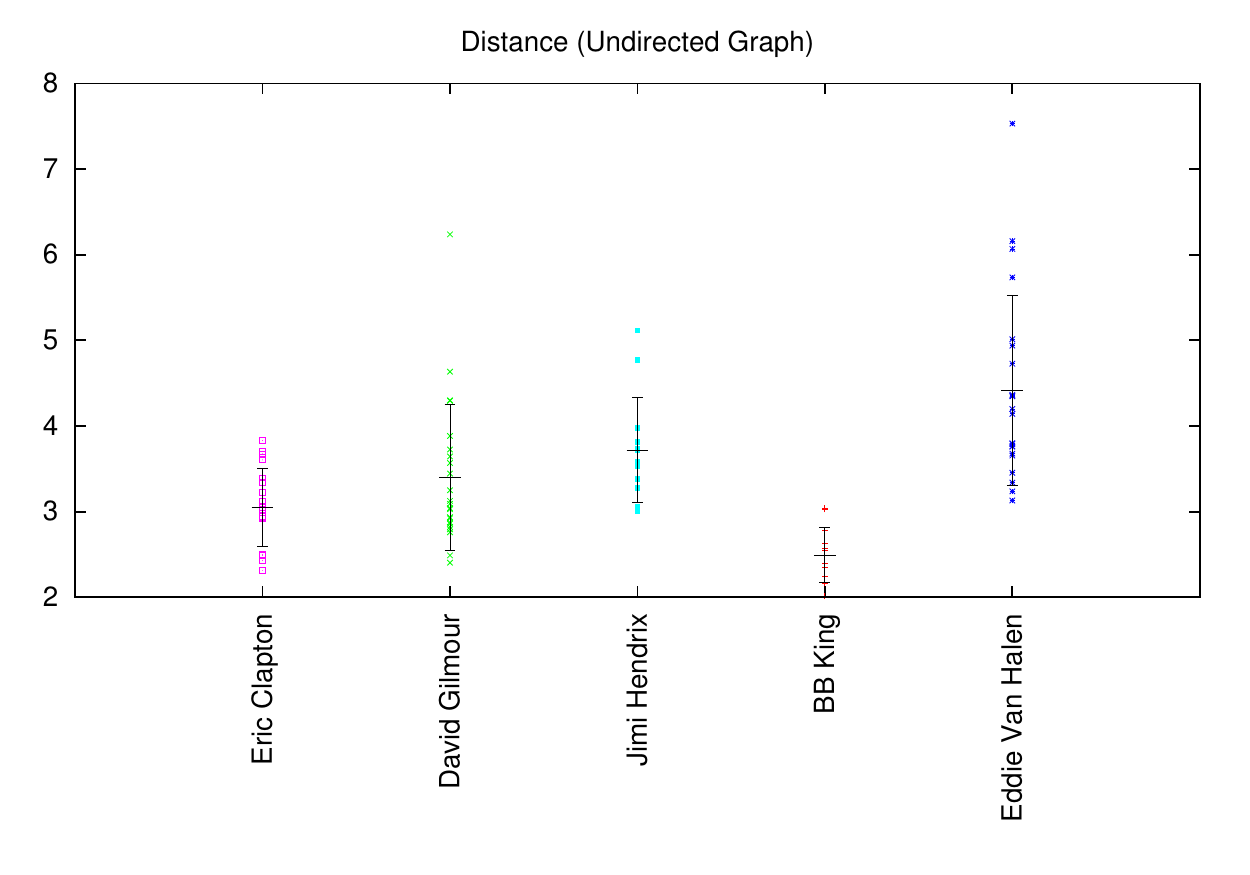}
   \vspace{-.5cm}
   \caption{Distance (undirected network)}
   \label{fig:undirectedDistance}
   \vspace{-.3cm}
\end{figure}

Figure \ref{fig:undirectedDistance} shows results for the considered performers. Outcomes confirm the claims reported for the degrees.
In general, classic blues performers (B.B.~King) have lower distances with respect to rock virtuoso performers (Eddie van Halen). 
Indeed, this outcome is in complete accordance with the common opinions of music experts.

\subsection{Clustering coefficient}

The clustering coefficient is a measure assessing how much nodes in a graph tend to cluster together.
It measures to what extent friends of a node are friends of one another too.
When two connected nodes have a common neighbor, this triplet of nodes form a triangle. The clustering coefficient is defined as 
$$C = \frac{3 \times \text{number of triangles in the network}}{\text{number of connected triplets of nodes}}$$
where a ``connected triple'' consists of a single node with links reaching a pair of other nodes; or, in other words, a connected triple is a set of three nodes connected by (at least) two links \cite{newmanHandbook}. A triangle of nodes forms three connected triplets, thus explaining the factor of three in the formula.

In this context, a triangle of nodes means that the performer played in a solo the corresponding three notes in sequence, but in different orders. In other words, the presence of many triangles (i.e.~high clustering coefficient) indicates that the performer is used to play different clusters of notes in a whatever order. Conversely, a low clustering coefficient might indicate that the performer prefers following a specific melodic line with a specific order on triplets.

\begin{figure}[htbp]
   \centering
   \includegraphics[width=.8\linewidth]{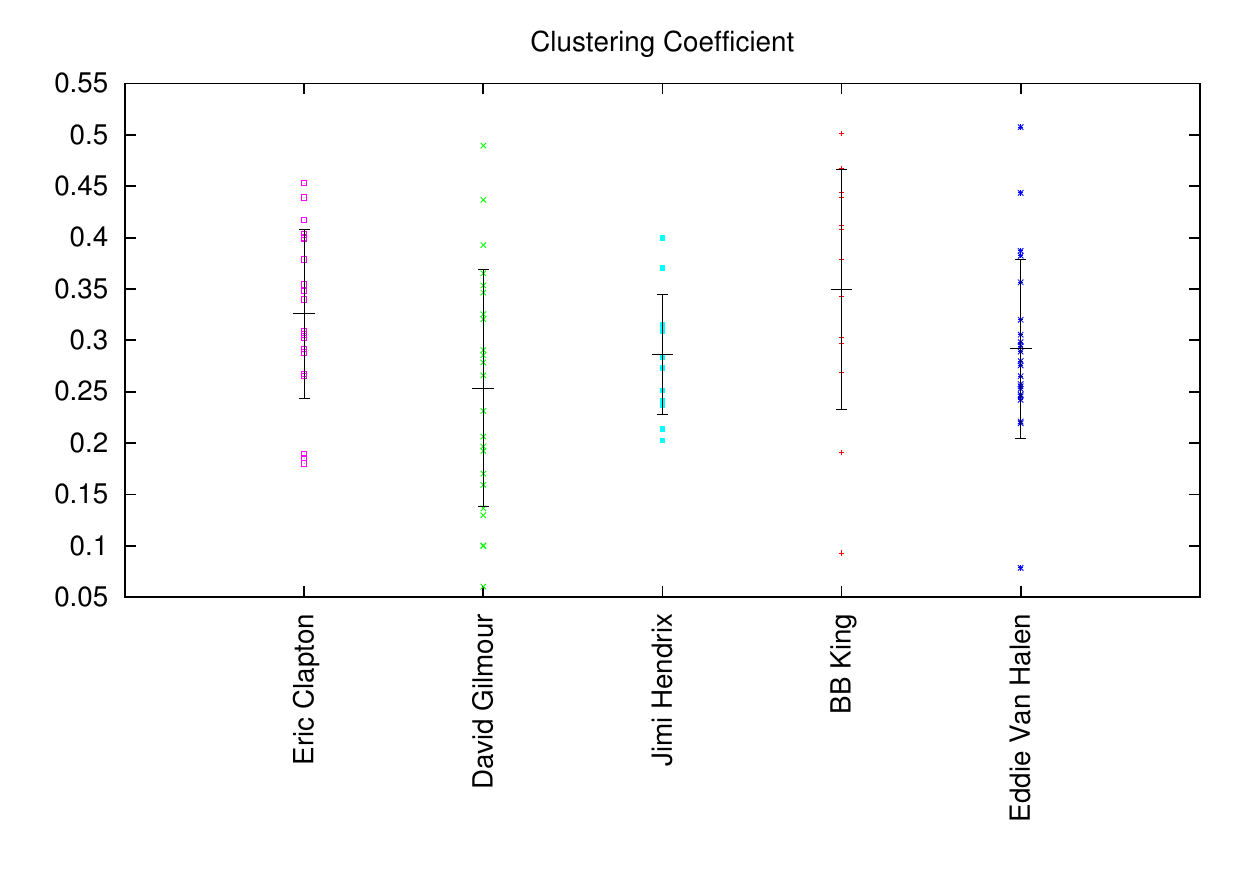}
      \vspace{-.5cm}
   \caption{Clustering coefficient}
   \label{fig:clusteringCoefficient}
   \vspace{-.3cm}
\end{figure}

Figure \ref{fig:clusteringCoefficient} shows the measured clustering coefficient for the considered performers. In this case, there no big differences among performers; David Gilmour has a lower average clustering coefficient, but with a high standard deviation.

\begin{table*}[th]
\centering
\caption{Small world property: comparison between solo networks and corresponding random graphs.}
\label{tab:sw}
\small
\begin{tabular}{|| l || c | c | c | c ||}
  \hline			
  \hline			
  Solo & clus coeff & clus coeff (RG) & avg distance & avg distance (RG)  \\
  \hline  
  \hline			
  B.B.~King -- Rock me baby & 0.41 & 0.11	& 2.17 & 3.26 \\
  \hline			
  D.~Gilmour (Pink Floyd) -- Comfortably numb (1st solo) & 0.06 &	0.03 &	4.30 &	4.03 \\
  \hline			
  E.~Clapton (Cream) -- Crossroads (2nd solo) & 0.40 &	0.04 &	3.68 &	4.29 \\
  \hline			
  J.~Hendrix -- Red House & 0.24 &	0.02 &	3.37 &	5.00 \\
  \hline  
  \hline			
\end{tabular}
\end{table*}

\subsection{Is there any small world property?}


Small world networks are networks that are ``highly clustered, like regular lattices, yet have small characteristic path lengths, like random graphs'' 
\cite{newmanHandbook}.
In a small world, most nodes are not linked with each other, but most nodes can be reached from every other by a small number of hops. Indeed, in a small-world network the typical distance between two randomly chosen nodes grows proportionally to the logarithm of the number of nodes.
Given a network, it is possible to verify if it is a small world, by comparing it with a random graph of the same size (i.e.~a network with links randomly generated, based on a simple probabilistic model).
Random graphs exhibit a small average distance among nodes (varying typically as the logarithm of the number of nodes) along with a small clustering coefficient.
In practice, one can assess whether a network has a small average distance as for a random graph, but a significantly higher clustering coefficient. In this case, the network is a small world.

Table \ref{tab:sw} reports results of this comparison for the four networks (solos) shown in Figure \ref{fig:someNets}. For each solo, column ``clus coeff'' of the table reports the clustering coefficient measured for the corresponding network; column ``clus coeff (RG)'' shows the clustering coefficient for a random graph of the same size; column ``avg distance'' provides the average distance among nodes in the networks (i.e.~the average shortest path); finally, column ``avg distance (RG)'' reports the average distance for a random graph of the same size.

It is possible to observe that two solos can be classified as small worlds, namely ``Crossroads'' by E.~Clapton and ``Red House'' by J.~Hendrix. In fact, these two solos have a clustering coefficient significantly higher than their corresponding random graphs, and the average distances are lower but comparable to those obtained for random graphs.
As to ``Rock me baby'', by B.B.~King, its clustering coefficient is lower than that of its corresponding random graph (and its average distance is lower as well), but this difference is not evident as for the other two solos mentioned above.
Finally, a small world property is not evident for ``Comfortably numb (1st solo)'' by D.~Gilmour. Indeed, this solo is a (touching) linear melody, without an intricate structure, and this is evident by the pictorial representation of the network reported in Figure \ref{fig:confNumb}.

\section{Conclusions}
\label{sec:conc}

In this paper, a novel approach has been presented to model melodies (and music pieces in general) as networks. 
We have analyzed a database of solos 
and extracted some main me\-trics. Outcomes are in complete accordance with the common opinions of music experts.
We argue that the use of a mathematical modeling of a solo (or a music track in general) provides a general and compact way to analyze music.

Results of this study can be exploited to build novel media applications related to music classification and categorization. 
Moreover, one might employ this framework as a main tool during the automatic generation of music. If we are able to harness the main characteristics of a musician, it would be possible to combine this approach to some artificial intelligence tool and generate, for example, a solo ``à la'' Miles Davis. 
As an example, these results suggest that 
a ``bluesy'' solo should have a high clustering coefficient and low distance;
a ``modern rock'' solo should have high distance and an average degree lower than other genres,
while a ``melodic'' solo should have a simpler structure.
This might have interesting applications in music 
didactics, multimedia entertainment, and digital music generation.

\small{

}

\end{document}